\documentclass[ aip, jap, amsmath,amssymb,
reprint,%
]{revtex4-1}

\usepackage{graphicx}
\usepackage{dcolumn}
\usepackage{bm}

\usepackage[utf8]{inputenc}
\usepackage[T1]{fontenc}
\usepackage{mathptmx}
\usepackage{soul}
\usepackage[colorlinks = true,
            linkcolor = red,
            urlcolor  = blue,
            citecolor = blue,
            anchorcolor = blue]{hyperref}
\usepackage{xcolor}
\usepackage{epstopdf}
\usepackage{siunitx}
\begin{document}


\title[Sample title]{Proof-of-concept thermoelectric oxygen sensor exploiting oxygen mobility of $GdBaCo_2O_{5+\delta}$}

\author{Soumya Biswas}
\author{Madhujith M K}%

\author{Vinayak B Kamble}
\email{kbvinayak@iisertvm.ac.in}
\affiliation{%
School of physics, Indian Institute of Science Education and Research Thiruvananthapuram, India 695551.
}%


\date{\today}

\begin{abstract}
 In this paper we demonstrate a proof-of-concept oxygen sensor based on the thermoelectric principle using polycrystalline $GdBaCo_2O_{5+\delta}$ where $0.45<\delta<0.55$ (GDCO). The lattice oxygen in layered double perovskite oxides is highly susceptible to the ambient oxygen partial pressure. The as-synthesized GDCO sample processed in ambient conditions shows pure orthorhombic ($P_{mmm}$ space group) phase and a $\delta$-value close to 0.5 as confirmed from X-ray diffraction reitveld refinement. The X-ray photoelectron spectroscopy shows significant $Co^{3+}$ oxidation state in non-octahedral sites in addition to $Co^{3+}$ and $Co^{4+}$ in octahedral sites. The insulator-to-metal transition (MIT) is observed at nearly 340 K as seen in electrical resistivity and seebeck coefficient measurements. The seebeck coefficient shows a large change of about (10-13 $\mu$V/K) with time constant of $\mathtt{\sim 20}$ sec, at room temperature (300 K) when the gas ambience changes from 100\% oxygen to 100\% nitrogen and vice versa, under a constant temperature gradient of 1K. The response in seebeck is found to be particularly large below MIT. The diffusion of oxygen into the lattice leading to hole doping shows a large change in carrier concentration resulting in a large change in the seebeck coefficient in insulating state. On the other hand, due to insignificant increase in already large carrier concentration in metallic state the change in seebeck is minimal. Nevertheless, below MIT the response is fairly reproducible within stoichiometry $\delta$ = 0.5 $\pm$ 0.05. This principle shall be of significant utility to design the oxygen sensors which work at room temperature or even cryogenic temperatures. 
\end{abstract}

\maketitle

\section{Introduction}
The perovskites are very important class of materials with diverse range of properties for potential applications like sensors\cite{Harwell2020, fergus2007}, dielectrics\cite{davies2008}, thermoelectrics\cite{Sukanti2017, maiti2019}, magnetic materials\cite{pardo2009}, multiferroics\cite{Tokunaga2009}, supercondutivity\cite{Manju2020} etc. The cobaltate based double perovskites particularly are complex crystals and often strongly correlated systems\cite{ikeda2016, pardo2009, Vasala2010, taskin2005, taskin2006}. The interest in these compounds arises due to spin and charge degrees of freedom owing to strong correlation among entities like oxygen content(carrier doping), rare earth ion radius and nonstoichiometry of central B-site ions (charge and spin)\cite{pardo2009, Vasala2010, taskin2005, taskin2006}. Particularly, in $AA^{\prime}B_2O_6$ type structure, the non-stoichiometry of central (B site) Co ions introduces various exotic properties like electron-hole asymmetry\cite{taskin2005PRL}, metal-insulator transitions\cite{frontera2002}, magnetic phase transitions\cite{taskin2005}, oxygen ion mobility etc in $ReBaCo_2O_6$. Besides, doping at Co site or varying the oxygen content of the lattice, induces either hole doping or electron doping in the system\cite{taskin2005, taskin2006}.
\begin{figure}[b]
\includegraphics[width=8.5cm]{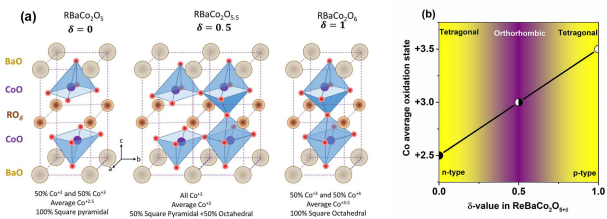}
\caption{The schematic diagram showing (a) the crystal structure and (b) phase diagram of the $GdBaCo_2O_{5+\delta}$, with average Co oxidation state for  $\delta$ =0, 0.5 and 1.0 showing tetragonal, orthorhombic and tetragonal unit cells respectively. }
\label{Fig1}
\end{figure} 

As shown in Figure \ref{Fig1}, the crystal structure of double perovskite type  $RBaCo_2O_{5+\delta}$ (RBCO)(where R is a rare earth element) consists of a sequence of metal oxide layers like CoO$_2$-BaO-CoO$_2-$ReO$_\delta$ periodic layers along the $\bar{c}$-axis. The $\delta$-value, i.e. oxygen content of the lattice mainly depends on the valency of Co, which can be 2+, 3+ or 4+. In a regular double perovskite $AA^{\prime}B_2O_6$ lattice i.e. $\delta$ =1, all Co ions have 50:50 3+:4+ i.e. all octahedra(O) of $CoO_6$; where 50\% Co ions have 3+ oxidation state and 50\% Co ions have 4+ oxidation state. While $\delta$ = 0, all Co ions have 50:50 2+:3+ (i.e. all square pyramids (P) of $CoO_5$; where 50\% Co ions have 2+ oxidation state and 50\% Co ions have 3+ oxidation state. However, at $\delta$= 0.5, all the Co are in 3+ state, i.e. there is exact equal contributions of $CoO_5$ – square pyramids and $CoO_6$ – octahedra. This causes the crystal structure to change from tetragonal ($\delta$ = 0) to orthrhombic ($\delta$ = 0.5) to again tetragonal ($\delta$ = 1) as shown in Fig. \ref{Fig1}(b). This results in varying physical properties such as, magnetic signatures of the composition varies due to either high spin, low spin or intermediate spin state of the Co ion\cite{takubo2006, taskin2005}. Further, the oxygen ions incorporated play a vital role in the properties of layered cobaltates\cite{taskin2005, taskin2006, liu2011}. The weak bonding of oxygen in $ReO_\delta$ layer enhances the oxygen diffusivity and hence increases the mobility of oxygen through the surface\cite{hermet2010, tsvetkov2014}. This in-turn shows a large change in transport properties of this systems as it is proposed that the equilibrium $\delta$ value at a given temperature linearly varies with the logarithm of the oxygen partial pressure (except for $\delta$ = 0)\cite{taskin2005}. Because of this, the double perovskites (AA$^\prime$B$_2$O$_{5+\delta}$ ) systems are rich in phases, like antiferromagnetic insulator, ferromagnetic insulator, ferromagnetic metal, paramagnetic metal etc\cite{taskin2005, ahmed2017}. They distinctly show insulator to metal transition (for intermediate $\delta$ values) near room temperature (~340 K for GBCO) and an order-disorder phase transition at relatively high temperature, (723 K for GBCO)\cite{liu2011}. This order-disorder phase transition is particularly characterized by a rearrangement of oxygen and its vacancies in the lattice. It may result in one-dimensional ordered oxygen vacancies along $\bar{a}$-axis and two-dimensional distributed vacancies in ReO$_\delta$ plane at low and high temperature respectively. This eases mobility of oxygen causing a change in the electronic transport of the materials, which can efficiently be utilised for monitoring oxygen levels for various applications like oxygen storage or fuel cells etc\cite{tarancon2008}. The beauty of layered rare earth cobaltates is that the change in annealing conditions  can controllably alter the oxygen concentration over a wide range in the $ReO_\delta$ planes. Here, the size of rare earth ion $Re^{+3}$ governs the equilibrium $\delta$ value at given temperature. Thus, because of the intermediate size of $Gd^{+3}$ ion among the lanthanide series, it was chosen as it allows a wide range of oxygen concentration. Besides, the air synthesized samples of Gd cobaltate show the $\delta$ value of near 0.5 which allows possibility of both, electron as well as hole doping and thus either decreasing or increasing $\delta$ value. 

The most successful oxygen sensors utilises wide bandgap metal oxide semiconductors like ZrO$_2$ which works on either potentiometric or amperiometric principles\cite{ramamoorthy2003, haaland1977}. These sensors require a constant reference on one side which is compared with the ambient oxygen level. Further to limit the current a small pin hole is used which is a tedious task to achive. Moreover, most of these sensors (either conductometric or amperometric or potentiometric) operate at much higher temperature i.e. 700 K and above which is practically difficult to maintain as it requires a power hungry heater. Besides, they limited applicability for applications like high altitude air breathability where the temperatures are fairly low. 

In this paper, a simple, proof of concept oxygen sensor has been proposed which exploits the sensitivity of thermoemf to the oxygen partial pressure and operates at room temperatures or even lower. The response is measured in change in open circuit voltage for a constant temperature difference. The response is large below the metal insulator transition temperature of GBCO which is 340 K. Thus, crucial role of mobile lattice oxygen in these double perovskite structures is studied thoroughly and the potential device is proposed that allows monitoring of oxygen, with reliable sensitivity using a more straightforward thermoemf measurement for use in monitoring the ambient oxygen breathability and control.

\section{Experimental}
\subsection{Synthesis}
The $GdBaCo_2O_{5+\delta}$ was prepared by solid-state synthesis technique. In a typical synthesis, gadolinium oxides (from Sigma Aldrich, Germany) was preheated at 800 $^o$C for 12 hrs to remove any absorbed volatilities from it. The stoichiometric amounts of the rare earth oxides were mixed with $BaCO_3$ (from Sigma Aldrich, Germany) and $Co(NO_3)_2.6H_2O$ (from Sigma Aldrich, Germany) to get an amount of 10 gm of  $GdBaCo_2O_{5+\delta}$. The composition was ground for about 2 hours in ethanol medium and was heated at 850 $^o$C for 24 hrs. Further, it was cooled naturally and grinding was repeated for 2 hrs. Subsequently, it was heated at 1100 $^o$C for 24 hrs.

\subsection{Material characterizations}
The X-ray diffraction (XRD) was recorded to confirm its crystal structure, phase formation and to estimate the crystallite size. The high resolution XRD data was recorded in 2$\theta$ range 5 to 90 degrees at a step size of 0.02 degrees at room temperature with sufficient exposure time to get good quality data on Philiphs panalytical X'pert pro diffractometer equipped with an accelerated detector. In order to reduce the grain size, planetary ball milling was done on Fritsch system (model Pulverisette 7) for 12 hrs for each sample, and again the XRD was taken after ball milling to compare the crystallinity.
The samples were pelletized by cold pressing and was heated at 1100 $^o$C for 24 hrs. To check the crystallinity and purity, XRD was repeated. The X-ray Photoelectron Spectroscopy (XPS) measurements were undertaken to verify the stoichiometry of each element in the compound. The electrical resistivity of sintered pellet was studied in van der Pauw geometry in an in-house built system as a function of temperature in vacuum ($10^{-2}$ torr). Keithley 2401 source meter was used to supply current and Keithley 2700 digital multimeter was used to measure the potential drop in four probes. The system was interface to personal computer for data logging.

\begin{figure}[b]
\center
\includegraphics[width=8.5cm]{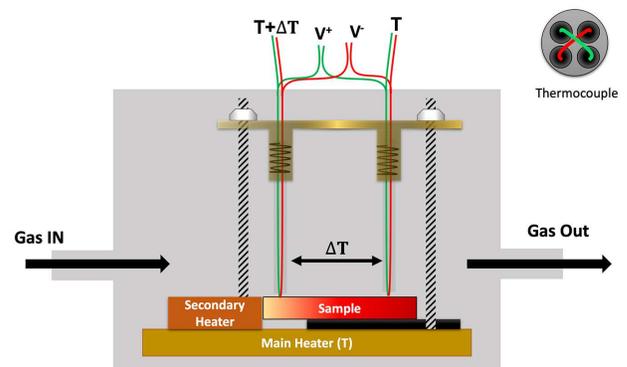}
\caption{The schematic diagram of the thermoelectric gas sensor system with the cartoon of thermocouple junction made by crossing the wires shown on top-right corner.}
\label{Fig2}
\end{figure}
\subsection{Thermoelectric gas sensing measurements}
An in-house seebeck based gas sensing system was developed to carry out the measurements. Figure \ref{Fig2} shows the schematic diagram of the gas sensing system. In a stainless steel chamber the sample was placed on a heater stage, with a thin alumina substrate below such that some part of the sample is floating. Thus, naturally there exists a small temperature gradient which is exploited to generate and measure thermoemf. The thermocouples are mounted on a spring base to ensure proper pressure at point of contact. In order to sweep the temperature gradient form negative to positive (for seebeck measurements) a secondary heater was placed touching the floating end of the sample. This heater was controlled using a Lakeshore 336 temperature controller for precise control. 
The two thermocouples used for measuring the temperatures and open circuit voltage are made by crossing two alumel-chromel wires (36 guage) through four bore alumina ceramics as shown in top-right in Fig \ref{Fig2}. This design allows to make a precise junction which is in contact with sample and avoids cold finger effect due to thinner thermocouple junction\cite{iwanaga2011}. Besides, the negligible thermal mass of the thermocouple ensured high sensitivity avoiding the time delay to attain  thermal equilibrium.

\section{Results and Discussion}
\subsection{Structural studies using X-ray Diffraction}
\begin{figure}[t]
\includegraphics[width=8cm]{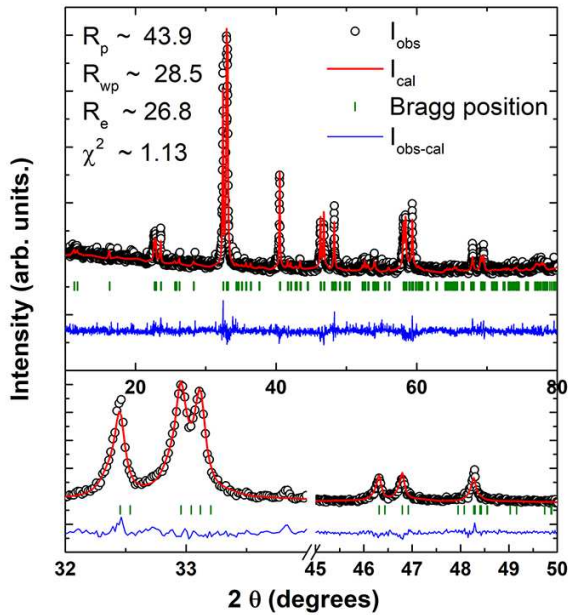}
\caption{Rietveld analysis of X-ray diffraction patterns of as prepared $GdBaCo_2O_{5.5}$  with $\delta$=0.5. The bottom panel shows the magnified view peaks showing splitting in (0kl) and (h0l) reflections}
\label{Fig3}
\end{figure}

The XRD patterns were collected for as synthesized powder as well as after ball milling. All the samples show the pure phase nature of double perovskites synthesized (see supplementary information section Fig S1). The XRD peaks are considerably broadened due to smaller crystallite size as a result of ball milling. The crystallite size was calculated from the full width at half maxima of the diffraction peaks using Scherrer formula shown in the equation \ref{scherrer},
\begin{equation}
t= \frac{0.9\lambda}{ \sqrt{\beta^2-\omega^2} \times cos \theta}
\label{scherrer}
\end{equation}

 where, 2$\theta$ is the Bragg angle, $\omega$ is the full width at half maxima and $\lambda$ is the wavelength of the incident x-ray (Cu k$_\alpha$ = 1.5418 $\si{\angstrom}$). The crystallite sizes computed from the XRD patterns are 130 and 95 nm before and after ball milling respectively. Thus, the crystallite size(t) was significantly reduced after ball milling. This size reduction was undertaken to improve the sinterability of the powders so that high densities are obtained to ensure high electrical conductivity of the sample. Since after ball milling the sample prepared for measurement by palletization. 

The X-ray diffraction (XRD) pattern of the GDCO powder sample is shown in Fig. \ref{Fig3} It was analysed by  the rietveld refinement method to evaluate the crystal structure using Fullprof (version 5.6) software package. As seen from Fig \ref{Fig3}, the pattern showed excellent goodness of fit (GoF) for observed and calculated data for the orthorhombic (Pmmm) structure corresponding to the $\delta$ value of 0.5 i.e $GdBaCo_2O_{5.5}$. The reduced $\chi^2$ value was 1.13 as shown in figure which signifies a very close fit. The lattice parameters obtained after refinement were a = 3.8793 $\si{\angstrom}$, b = 7.8350 $\si{\angstrom}$   and c = 7.5349 $\si{\angstrom}$ . These values agrees well with those reported in literature for orthorhombic phase having $\delta$ between 0.45 and 0.55\cite{taskin2005PRL}.

As mentioned earlier, the double perovskites can have variable lattice oxygen contents and this is primarily governed by the processing conditions. This large variation in oxygen site occupancy results in structural phase transitions as a function of oxygen content, precise $0 \leq \delta \leq 1 $ values. When the structure has low oxygen contents i.e. $\delta<0.5$ The structure is tetragonal. i.e. two of the three crystallographic axes are degenerate and hence only one single peak is obtained for degenerate (0kl) and (h0l) reflections for same value of $l$ index. On the other hand, when $\delta$ value increases and is close to 0.5, the structure becomes orthorhombic (space group $Pmmm$) as systematically one oxygen from every alternate $CoO_6$ octahedra is missing. Thus, (0kl) and (h0l) are no more the same spacing. In fact the lattice constant of $\bar{b}$ -axis is doubled as shown in Fig \ref{Fig1} (however, the second order peak of (0 k/2 l) still appears close to (h0l). Hence, (0kl) and (h0l) gives two different bragg peaks depending on the difference in $\bar{a}$ and $\bar{b}$ lattice parameter. Further increase in $\delta$ makes the system again tetragonal and hence(0kl) and (h0l) again becomes same for same value of $l$. Thus, in our samples it was observed that GBCO shows two splitted peaks (see supplementary information Fig S1). Based on this one may conclude that GBCO is completely  orthorhombic thus, their $\delta$ value is close to 0.5, maybe just lesser or little more. 

\subsection{X-ray Photoelectron spectroscopy}
\begin{figure*}
\center
\includegraphics[width= 16cm]{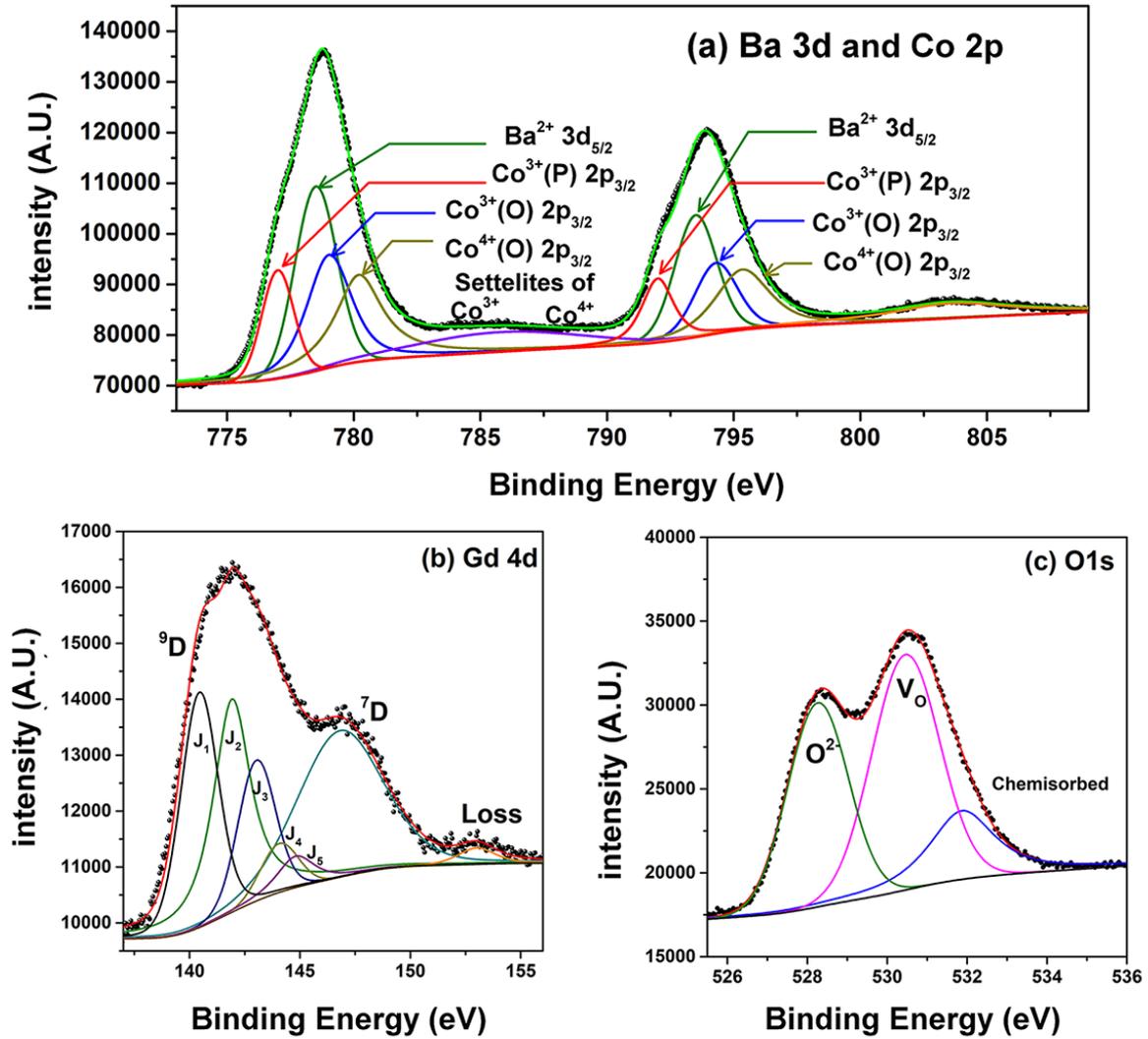}
\caption{The X-ray Photoelectron spectra of (a) Ba $3d$ and Co $2p$ (b) Gd $4d$ and (c) O $1s$ core levels.}
\label{Fig4}
\end{figure*}

The X-ray Photoelectron Spectra of the as-synthesized sample was studied. The core level spectra of Ba $3d$, Co $2p$ and Gd $4d$ and O $1s$ are shown in \ref{Fig4}. Although there are a number of reports in the literature showing X-ray photoelectron spectra (XPS) of double perovskites, most of them discuss the spectra qualitatively and do not show any deconvolution. In this work, we seek to analyse the XPS spectra of these materials for the first time to the best of our knowledge and establish a correlation with observed electronic transport properties. The analysis of XPS spectrum of these systems is rather challenging due to several complexities like, variable oxidation state of Co and its overlap of $2p$ line with Ba $3d$; variation in the oxygen content in the system ($\delta$-value) and the inherent magnetic moments of certain rare earth elements like Gd. Thus, the XPS analysis of these systems is non-trivial and hence could be of great relevance for researchers interested in these 112 type double perovskite systems. Ideally, for Co 2p level has the highest photoionisation cross-section, However, since Co $2p$ and Ba $3d$ core levels are overlapping in binding energy, we have also investigated the next prominent core level spectra, i.e. Co $3p$ and Ba $4d$ respectively and resolved the spectrum.

Fig. \ref{Fig4}(a) shows the XPS spectrum of Co $2p$ and Ba $3d$ overlapping core levels. Here several peaks have been identified considering spin-orbit splitting of $3d$ ($l$=2) of Ba and $2p$ ($l$=1) of Co photo-emissions. The Ba $3d_{5/2}$ and $3d_{3/2}$  lines have been fixed with reference to those observed in similar systems such as by Maiti et.al.\cite{Maiti2009} and others \cite{fetisov2015}. Maiti et.al.\cite{Maiti2009} have shown that in case of $YBa_2Cu_3O_{6+\delta}$, which is also oxygen deficient perovskite system, the binding energy of Ba $3d$ line lies in between 777 and 778 eV for $\delta$ =0.5 $\pm$ 0.05.  The same has been confirmed by the spectra shown by Pramana et.al.\cite{pramana2018}. Thus we fixed the Ba 3d$_{5/2}$ line in the spectrum at this binding energy and tried to adjust the deconvolution for Co ions accordingly. The Ba $3d_{3/2}$ line was marked considering the spin-orbit splitting energy of Ba 3d line from the literature\cite{takubo2006} which marks same position for ba 3d 5/2 for $\delta$=0.5. 

Regarding fixing the oxidation state of Co ions is challenging here because of possibility of variable oxidation state and corresponding satellites. However, the presence of satellites itself can be effectively used for fixing the oxidation state of transition metal ions\cite{biesinger2011}. Thus, the two humps seen at 785 and 789 eV (shown by single broad peak) are ascribed to the satellite peaks of $Co^{3+}$ and $Co^{4+}$ oxidation states respectively\cite{biesinger2011}. As shown by Takubo et.al.\cite{takubo2006}, in case of Nd and Tb based rare earth cobaltate, the Co $2p_{3/2}$ is expected at 780 eV. This is the most intense and hence abundant oxidation state in the structure i.e. $Co^{3+}$ in octahedral (O) position for $\delta=0.5$ composition. Thus, the next oxidation state i.e. $Co^{4+}$ which is also in   Octahedral site is hardly 1 eV higher than its predecessor i.e.$Co^{3+}$(O). Besides there 2p$_{1/2}$ counterparts are included  at 794 and 795 eV respectively.

In addition to the peaks mentioned above, there is a shoulder seen to the left of the peak, which could neither be attributed to $Co^{3+}$(O) and $Co^{4+}$(O). Besides, it could also not be $Co^{2+}$ because it is unlikely to occupy an { GDCO ($\delta <$ 0.5) (a) at 333 K }octahedral site while having a 2+ charge. Even if it would, it would not have been so lower binding energy than $Co^{3+}$(O). Thus, this has been identified as the $Co^{3+}$ ions occupying a crystallographic site other than that of octahedral one which is square planar site (marked by P) in this case. Since the binding energy of photoelectron ejected from a particular atom is highly sensitive to its local chemical environment in addition to the nuclear charge. Thus, like chemical shift due to variable oxidation state of the emitting atoms, it is also found that the binding energies of photoemitted electrons are also slightly different for same element, having same oxidation states but different crystallographic sites\cite{reiche2000}. 

Fig. \ref{Fig4}(b) shows the Gd 4d core level of the XPS spectrum. Usually, in case of d orbital which has azimuthal quantum no $l=2$, the two spin orbit interactions expected are $j_1=l+s$ and $j_2=l-s$ where $s=1/2$ for the electron. which means only two lines of $j_1=5/2$, $j_2=3/2$ are expected. However, $Gd^{3+}$ being a $4f^7$ ion it shows a strong coulomb and exchange interaction between 4d and 4f electrons as presented in some of the literature\cite{lademan1996, talik2016} in spite of the spin integrated collection of photoelectrons than spin resolved spectra. The two intense lines corresponds to $^9D$ and $^7D$ final states which show spin-orbit interaction energy of nearly 6 eV. The spin parallel state $^9D$ shows finer spillting due to spin spin interaction, while the anti-parallel state $^7D$ could not be resolved.  These levels further split into 6 depending on separate $m_j$ values due to strong $4d-4f$ coupling of Gd ion. The broad peak at 154 eV is ascribed to energy loss feature. 

The oxygen $1s$ spectrum shows three contributions, which have been identified as the lattice oxygen peak (at 528.5 eV), the peak corresponding to lattice oxygen vacancies (530.5 eV) and the surface chemisorbed oxygen from atmosphere at about 532.5 eV. These accreditation are in good agreement with those of similar systems \cite{takubo2006}.

\subsection{Transport Measurements}
\subsubsection{Electrical resistivity}
\begin{figure}
\includegraphics[width= 8cm]{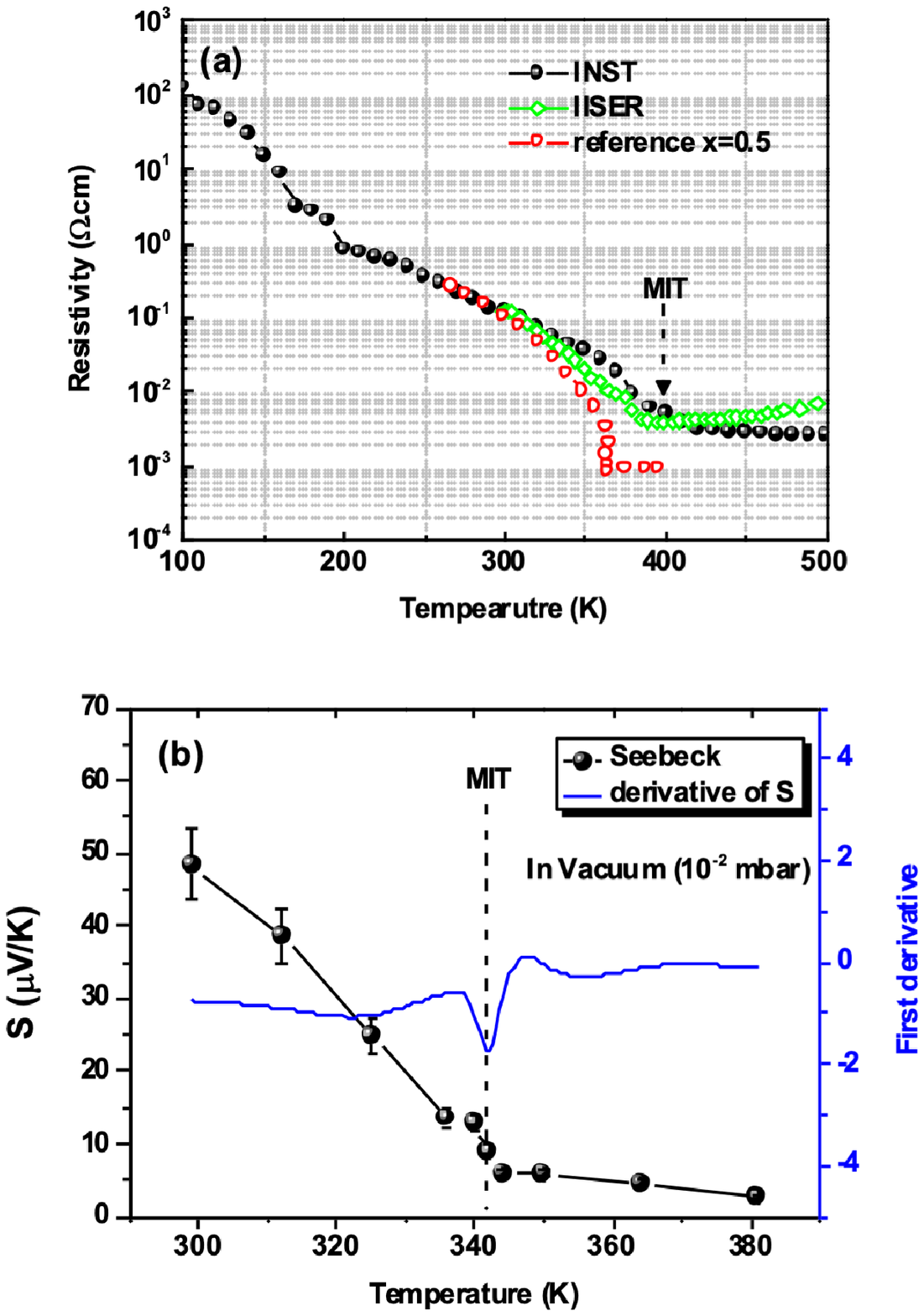}
\caption{(a) The variation of electrical resistivity of GDCO sample measured in two different systems and compared with that of the reported data. (b) The seebeck coeffocient measurement on the same sample across the metal insulator transition marked by first derivative of seebeck with temperature.}
\label{Fig5}
\end{figure}

The electrical resistivity and seebeck coefficient of the samples were measured in an in house built system. The measurement utilized Van der Paw method of four probes using the equation \ref{Resistivity_Eq}. The system was calibrated using Ni metal foil as standard. A constant current (10 mA to 1A) is applied and the voltage drop (V) has been measured as a function of temperature (T) of all the pellet samples. 
\begin{equation}
\rho = \frac{\pi R }{ln 2}\times d
\label{Resistivity_Eq}
\end{equation}
where $\rho$ is the electrical resistivity, R is the sheet resistance and d is the sample thickness.

Fig \ref{Fig5} shows the resistivity behaviour of as prepared sample in vacuum from 100 to 500K. in the lower temperature range, the resistivity shows a  drop until 330 K, after which it shows a sudden drop by  more than one order of magnitude. beyond which the resistivity shows a small positive slope. The same sample was measured in two different systems (marked as IISER and INST) to confirm the measurement data. Also, the data from Taskin et.al.\cite{taskin2006} has been shown for comparison and matching of delta value reference. From Figure \ref{Fig5}(a) it is clear that the data obtained matched very well with that of Taskin et.al.\cite{taskin2006}  for $\delta$ =0.5. The small discrepancy in temperature of transition could be due tothe fact that the data of Taskin et.al.\cite{taskin2006}  is measured on single crystals while that of this study is on a polycrystalline pellet sample. 

\begin{figure}
\includegraphics[width=8cm]{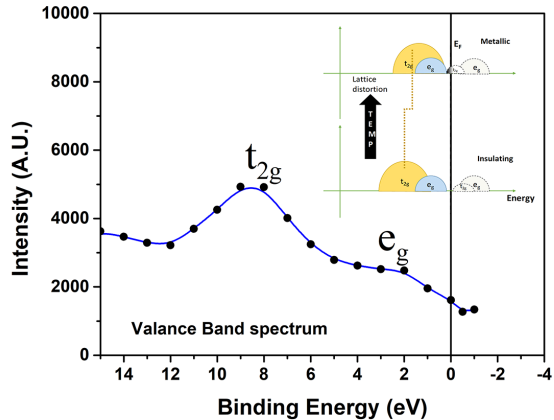}
\caption{The valance band region of X-ray photoelectron spectra showing the crystal filed energy levels of Co ion at room temperature. The inset cartoon on top right shows the change in electron occupancy at fermi level ($E_F$) when the system is heated beyond 350K, $CoO_6$ undergoes a lattice distortion changing the density of states at the $E_F$ from near zero (insulating) to non-zero (Metallic) behaviour.}
\label{Fig6}
\end{figure}

Tarancon et.al.\cite{tarancon2008} have performed high temperature XRD on GDCO and shown that at ~350 K, there happens a lattice distortion, i.e. lattice parameter $\bar{a}$ shrinks i..e corresponding spacing shows a shift towards higher Bragg angle; while lattice parameter $\bar{b}$ and $\bar{c}$ show a sudden elongation i.e. corresponding lattice spacing shows a shift towards lower Bragg angle. This lattice distortion causes the change in crystal Field splitting of the $CoO_6$  octahedra and thereby changing the position of $t_{2g}$ and $e_g$ orbitals relative to each other bring non-zero density of states at the Fermi level. The low binding energy region of XPS spectrum which essentially depicts the valance band is shown in Fig. \ref{Fig6} recorded at room temperature. This shows near flat region at $E_F$ i.e. zero binding energy. Because this is very close to the metal insulator transition(MIT) at about 350K, there may be a small non-zero electron occupancy even at room temperature. According to literature, the transition only occurs for compositions having a $\delta$ value of 0.5 and its vicinity i.e. 0.45 $< \delta <$ 0.65. If the delta value is above and below this range the resistivity shows an normal insulator like exponential decrease.
 
\begin{figure}[b]
\includegraphics[width= 8cm]{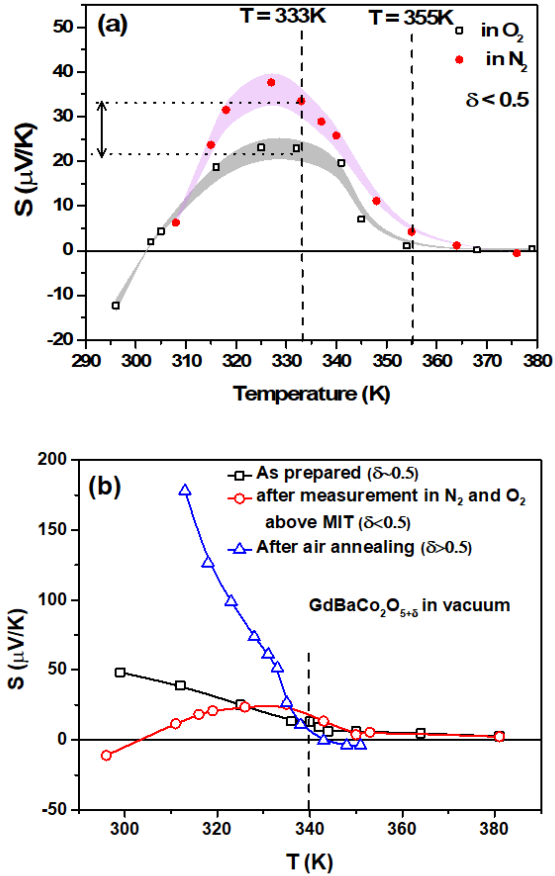}
\caption{(a) The variation of electrical resistivity of GDCO sample measured in two different systems and compared with that of the reported data. (b) The seebeck coefficient measurement on the same sample across the metal insulator transition marked by first derivative of seebeck with temperature.}
\label{Fig7}
\end{figure}

The signature of MIT is quite evident from the seebeck coefficient variation of the sample as shown in Fig. \ref{Fig5}(b). The seebeck coefficient is fairly large in the insulating state and decreases gradually as the resistivity decreases. Beyond the MIT temperature, S shows a marked change of slope and has very small value ($\mathtt{\sim 1 \mu V/K}$) marking a metallic behavior. 
The first derivative of S with respect to the temperature shows a clear minima which marks the MIT around 345K. The thermoemf behaviour of this system has been studied thoroughly an hence deduce the oxygen content of the lattice by corroborating the result with that of the literature. According to the detailed study conducted by Taskin et.al \cite{taskin2005, taskin2006}, the maximum temperature of MIT is obtained for $\delta=0.5$ and the temperature decreases as the value $\delta$ deviates from this value on either side. Moreover, the behavior of seebeck is the same for single crystals or polycrystalline bulk samples. However, from mere visual comparisons it can be deduced that the Seebeck values obtained here for as-prepared sample matches with that of the $\delta$ of 0.497, rather in between to that of 0.495 to 0.501. 
\\
According to literature, $Co^{3+}$ in octahedral coordination has a low spin (LS, $t_{2g}^6e_g^0$, S=0) state whereas in square pyramidal coordination same $Co^{3+}$ has intermediate spin (IS, $t_{2g}^5e_g^1$, S=1). On the other hand, when the system is doped with electrons, i.e. $Co^{2+}$ which occupies square pyramidal coordination has high spin (IS, $t_{2g}^5e_g^2$, S=3/2)and doped with holes i.e. $Co^{4+}$ which goes in octahedral coordination has low spin  ($t_{2g}^5e_g^0$, S=1/2). Thus, although the change is $\delta$ upon incorporation or removal of oxygen is symmetric about $\delta$=0.5, the transport properties are not symmetric. \cite{taskin2005, taskin2005PRL, taskin2006}. Thus, the system is more conducting for hole doping (i.e. $\delta >0.5$) and rather insulating for electron doping ($\delta<0.5$). This is due to the fact that the transport of HS electrons $Co^{2+}$ is rather suppressed due to spin blockade\cite{taskin2005PRL} in the background of $Co^{3+}$ which has LS. This is the reason, the seebeck coefficient of the system shows asymmetry in spite of divergence at 0.5 $\delta$ value. the seebeck coefficient for $\delta =0.5$ composition is rather negative at low temperature (<250K), which becomes large positive even with small change in $\delta$ i.e. 0.501. The  $\delta$ values Fig \ref{Fig7}(b) shows good correlation with these arguments. 

Thus, when the sample ambience changes from oxygen rich to oxygen lean, the sample seebeck value shows a large change as compared to its resistivity, which is is nearly same across the $\delta$ value. Here we measured the change in the seebeck coefficient by changing the atmosphere from $N_2$ to $O_2$ and back to see the change in the potential difference for a fixed temperature gradient across the the hot and cold end of the sample. The system is calibrated using Ni metal sample in the given temperature range. See Figure S5 and S6 in the supplementary information section for the Nickel calibration data and the data analysis of seebeck measurements. 
As seen in Fig \ref{Fig7}(a), the seebeck coefficient in oxygen atmosphere is lower than that of nitrogen atmosphere. This is due to diffusion of oxygen into the lattice, which doped the system with holes, makes it more conducting by oxidising some of $Co^{2+}$ to $Co^{3+}$, i.e. square pyramidal to octahedral coordination. However, this change is conspicuous only for high seebeck values. There maybe change in Seebeck when it is low, however, the resolution of measurement may be a limiting factor. Nevertheless, the ambient cycling at room temperature changes the oxygen stoichiometry. 

In order to estimate the change in seebeck upon ambient cycling for sample with high initial $\delta$ values, the sample was ground in mortar and pestle and annealed in ambient oxygen at 1000 $^oC$ for 2 hours and naturally cooled to room temperature. The sample so obtained showed a large seebeck value as shown in Figure \ref{Fig7}(b), which confirmed the value of $\delta$ to be greater than 0.5 (upon comparison with data in literature \cite{taskin2005}). The nominal MIT temperature value is shown by vertical dashed line in Figure \ref{Fig7}(b).

\begin{figure*}[]
\includegraphics[width= 15cm]{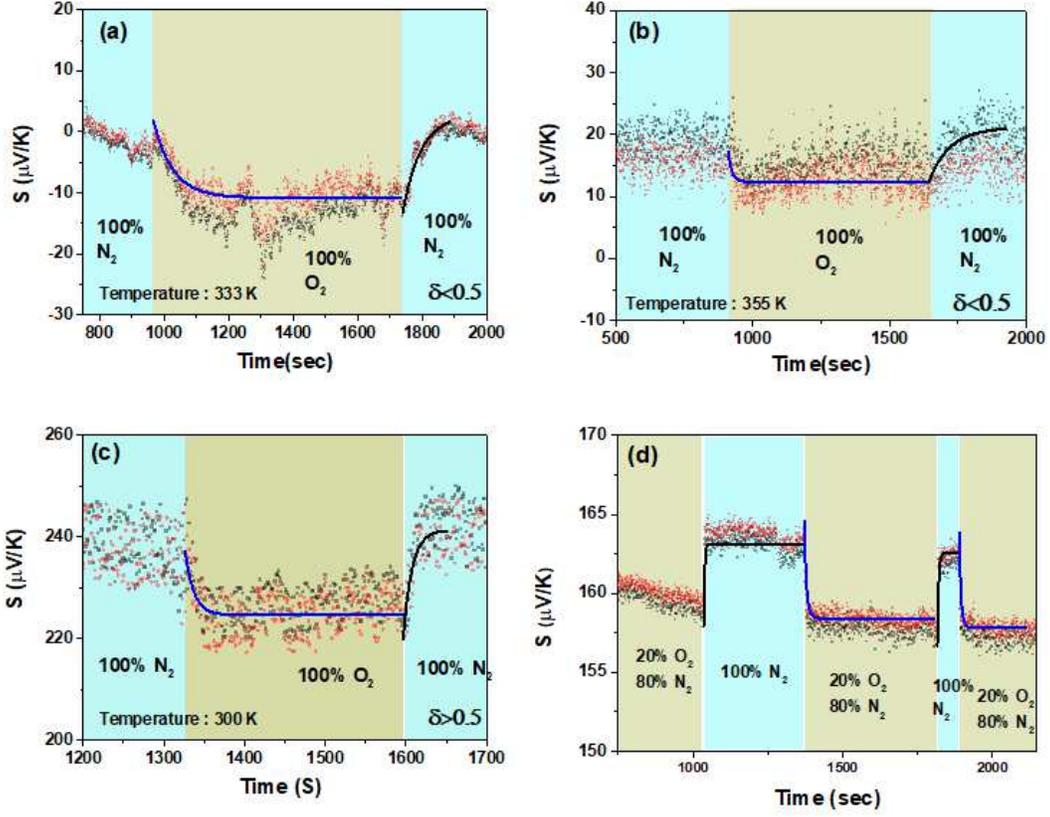}
\caption{ The response of seebeck coefficient of GDCO ($\delta <$ 0.5) (a) at 333 K and (b) at 355 K. (c) the response of seebeck coefficient of GDCO ($\delta >$ 0.5)  at 300 K for different gas ambience 100 \%$N_2$ to $O_2$ and back. (d) for 20\% $O_2$ and 100\% $N_2$ and back at 300 K.}
\label{Fig8}
\end{figure*}

The seebeck measurement data for sample which showed consistent results is shown in Figure \ref{Fig8}. As mentioned earlier after the first ambient cycling at room temperature the as synthesized sample ($\delta =0.5$) showed a small change in oxygen stoichiometry ($\delta <0.5$) as manifested in seebeck. However, the resulting oxygen stoichiometry was fairly consistent until it was re-annealed at high temperature to give a larger $\delta$ ($>0.5$) value. 
Thus, the Figure \ref{Fig8} (a) and (b) show the response in seebeck for the $\delta <0.5$ sample before and after MIT i.e. 340 K. As it is seen from Figure \ref{Fig8}(a) obtained at 333K shows a large and noticeable change in seebeck where as  \ref{Fig8}(b) obtained at 353 K does not.  On the other hand, the same trend is observed for sample showing large seebeck ( and $\delta >0.5$) at room temperature, 300 K shown in \ref{Fig8}(c). In either cases, for insulating state, a large change in seebeck has been observed (10-15 $\mu$V/K) and in metallic state there is no significant change in seeebeck. 

\begin{figure}[b]
\includegraphics[width= 8cm]{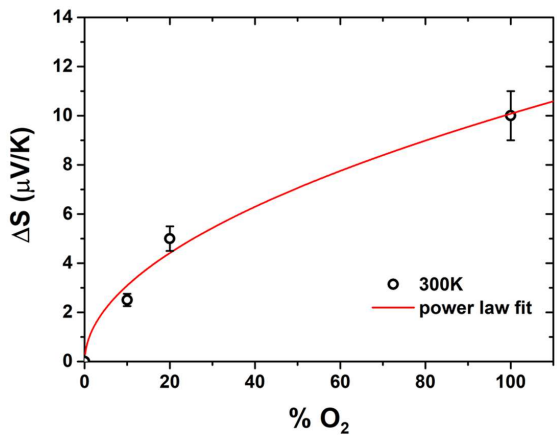}
\caption{The change in Seebeck coefficient ($\Delta S$) for different $O_2$ concentrations in \% with power law fit.}
\label{Fig9}
\end{figure}

We also measured the response in seebeck coefficient for fractional oxygen atmosphere such as ambient 20\% $O_2$ and balance 80\% $N_2$ to that of 100\% $N_2$. The result is shown in Figure \ref{Fig9}(a) wherein a clear sharp rise and fall is seen for change in atmosphere. Before measuring this response the chamber was first evacuated and then filled with 100\% $N_2$ gas and subsequently 20\% $O_2$ was introduced under flowing condition. similarly the change was recorded for 10\% $O_2$. Notably, the change in seebeck for a given concentration of $O_2$ was practically the same irrespective of the initial seebeck value. This means the change in seebeck coefficient was found to be independent of initial $\delta$ value except for where S it is really low. This change in S i.e. $\Delta S$ was plotted against the given $O_2$ concentration in percentage. Surprising the data was found to obey the power law of semiconductor gas sensors \cite{kamble2016SNB} as shown in Figure \ref{Fig9}. This implies that the response of the material is scaled as power of concentration of the gas.  Usually the response is measured in change in resistance, unlike change in seebeck in this case. The governing equation of the law is stated as-

\begin{equation}
\Delta S = AC^\beta
\label{power law}
\end{equation}
 where, A is a scaling prefactor and $\beta$ is the exponent which is governed by the gas-solid interaction. In this case the interaction is diffusion of oxygen through $ab$ plane which leads to change in seebeck coefficient of the system. This change could be due to the change in carrier concentration (n). As mentioned by Taskin et.al.\cite{taskin2005PRL}, even the small change in $\delta$ value of 0.001 i.e from 0.5 to 0.501, causes a large change in the carrier concentration of nearly $10^{19} cm^{-3}$. Because seebeck coefficient depends mainly on three quantities, carrier concentration(n), temperature (T) and effective mass (m*) i.e. slope of the density of states at $E_F$ as shown in equation below.
\begin{equation}
S = \frac{8\pi^2 k_B^2}{3eh^2} m* T \left[\frac{\pi}{3n}\right]^{2/3}
\end{equation} 

Thus, if the temperature of sample is held constant, only two quantities can change i.e. n and m*. However, it is unlikely that m* would change drastically giving such as large change in seebeck. moreover, when oxygen is diffused inside the lattice and occupies the vacant sites along the ordered oxygen vacancies in ab plane, it dopes the system with excess holes, making the system more conducting. thus, can produce significant change in n. This change is very much evident in S variation at lower temperature\cite{taskin2005}. Moreover as the $\delta$ values drops below 0.5 the seebeck diverges and shows first a low positive value ($\delta > 0.7$), large positive value ($0.7 > \delta > 0.5$), large negative value ($0.5 > \delta > 0.45 $) and moderately large negative value ($\delta < 0.45$). 

After the virgin test of change in ambience, it was found that the seebeck coefficient of the sample has shown a different trend which did not match to that of $\delta$ =0.5 as seen in Fig \ref{Fig7}(b). However, this trend was found to very reproducible ($\delta < 0.5$) at room temperature.  
The average oxygen content in the lattice ($\bar{\delta}$) has been found to show an exponential dependence with time and the time constant ($\tau$ ) depends on temperature\cite{taskin2005}.
\begin{equation}
\bar{\delta}= \delta_\infty - [\delta_\infty - \delta_0] e^{(-t/\tau)}
\label{time_constant}
\end{equation} 

where, $\delta_\infty $ is the oxygen content at equilibrium (t = $\infty$) and  $\delta_0 $ is the initial oxygen content (t=0).  This time constant is governed by the diffusion coefficient of oxygen into the lattice. As per the detailed kinetics studied by Taskin et.al.\cite{taskin2005}, as the temperature increases the diffusivity of oxygen increases and hence at higher temperature the oxygen content ($\delta$) lower at given oxygen partial pressure ($P_{O_2}$). However, after annealing at higher temperature for a long time (several hours) if the sample is naturally cooled to room temperature in a constant $P_{O_2}$, the lattice oxygen content rises as it cools and reaches the thermal equilibrium value of room temperature. Thus, the sample synthesized in atmospheric $P_{O_2} \mathtt{\sim 10^{-1}}$ bar shows a high $\delta$ value of near 0.5. The data obtained is fit with exponential functions according to equation \ref{time_constant} and the values of time constants $\tau$ is estimated for response as well as recovery. A very small value of about 10 sec is obtained for the time constant which is remarkably fast for a bulk diffusion phenomenon.

The kinetics of the response and recovery times are governed by the diffusion coefficient of oxygen at a given temperature which decides the time constant as shown in equation \ref{diffusivity}\cite{Sreya2019},

\begin{equation}
D=\frac{2 L^2}{\pi^2 \tau}
\label{diffusivity}
\end{equation}

where, $L^2$ is the area of surface available through which the diffusion takes place and $\tau$ is the time constant of diffusion for rise and recovery. 
According to Taskin et.al. \cite{taskin2005} the diffusivity (D) of oxygen in GDCO is 3 $\times 10^{-8} cm^2/sec$ at $250^oC$. Moreover, this diffusion being a thermally activated process, diffusivity value would be much lower at room temperature. However, using this value and typical surface area of our sample, i.e. 1 cm radius one gets the value of time constant as nearly $10^{+7} sec$  which is much much larger than that of found in our measurement. One of the reason could be the smaller crystallite size (100 nm) in our case as compared to that of Taskin et.al.(1 $\mu m$). Moreover, in our case since  the time constant is measured using transport measurements and the thermocouples being resting on the surface, we get a near instantaneous change in the surface voltage which saturates quickly.

The idea of thermoelectric gas sensor was first proposed by Retting and Moos \cite{rettig2007} which proposed viable design and demonstrated oxygen sensing abilities of $SrTi_{0.6}Fe_{0.4}O_3$ using thermoelectric prinicples at 750 $^oC$. Thus, one can prepare thin films of GDCO on a microheater platform which is inherently designed to give a small  (calibrated) temperature at a given heater current and thus measure its open circuit voltage. Higher the temperature gradient, higher is the voltage produced Subsequenty higher resolution is obtained between increasing oxygen concentrations. Similar thermoelectroc based sensor has been demonstrated by Masoumi et.al. \cite{masoumi2019} using ZnO for Volatile Organic Compound (VOC) sensing. However it also had much higher operating temperature (400 $^oC$) and very poor selectivity among VOCs. Nonetheless, similar power law dependance was observed. The primary mechanism identified for change in seebeck upon exposure to VOCs was change in the inter-grain barrier height due to carrier injection or trapping becuase of gas-surface interaction\cite{hossein2018}, unlike bulk oxygen diffusion in our case. 

\section{Conclusion}
Here we demonstrate that the polycrystalline bulk ceramic sample of $GdBaCo_2O_{5.5}$ (GDCO) prepared by solid state route showed an optimum oxygen content ($\delta$= 0.5) resulting in demonstration of metal insulator transition close to room temperature and an oxygen ambient dependant transport properties. We found that the Ba $3d$ and Co $2p$ core level regions overlap in XPS making this analysis challenging. However, with reference to literature on similar systems and transport properties, we have successfully analysed the XPS spectra to reveal significant disproportionation of Co ions into 3+ and 4+ along with $Ba^{2+}$ 3d photoemission. Moreover, interestingly, there exists a significant contribution of 3+ in non-octahedral coordination, which is characterised by rather lower binding energy than usual $Co^{n+}$ in octahedral site. This binding is lower due to absence of one high electron affinity oxide ion in square pyramidal geometry than that of octahedral. The transport properties reveal that the as prepared sample has $\delta$= 0.5 and shows an MIT at nearly 340 K which agrees well with literature. Further, the seebeck coefficient shows a step change at MIT. It also exhibits a large change upon change of oxygen content of the ambience. The small response and recovery time constant (nearly 10 sec) obtained at room temperature depict a surface sensitive nature of measurement method. Thus, this method of measuring open circuit voltage for a small temperature difference has an edge over the existing oxygen sensors involving very high temperature of operation ($\geq 500 ^oC$) and a reference oxygen level for comparison.

\begin{acknowledgements}
The authors are thankful the the funding revived from Science and Engineering Research Board (SERB) Govt. of India (Grant No EEQ/2018/000769). The authors are thankful to Dr Chandan Bera of INST Mohali, for electrical conductivity measurements at low temperature.  The authors are also thankful to Prof Arun M Umarji for his guidance about the system of double perovskites.
\end{acknowledgements}

The data that support the findings of this study are available from the corresponding author upon reasonable request.
\nocite{*}
\bibliography{aipsmp}
\end{document}